# ELECTRICALLY CONTROLLED LONG-DISTANCE SPIN TRANSPORT THROUGH AN ANTIFERROMAGNETIC INSULATOR

R. Lebrun[1,*], A. Ross[1,2,*], S. A. Bender[3], A. Qaiumzadeh[4], L. Baldrati[1], J. Cramer[1,2], A. Brataas[4], R. A. Duine[3,4,5] & M. Kläui[1,2,4]

[1] Institute for Physics, Johannes Gutenberg-University Mainz, 55099 Mainz, Germany
[2] Graduate School of Excellence Materials Science in Mainz, Staudingerweg 9, 55128, Mainz, Germany
[3] Utrecht University, Princetonplein 5, 3584 CC Utrecht, Netherlands
[4] Center for Quantum Spintronics, Department of Physics, Norwegian University of Science and Technology, NO-7491 Trondheim, Norway
[5] Department of Applied Physics, Eindhoven University of Technology, P.O. Box 513, 5600 MB Eindhoven, The Netherlands
* Both authors contributed equally to this work.

**Spintronics uses spins, the intrinsic angular momentum of electrons, as an alternative for the electron charge. Its long-term goal is in the development of beyond-Moore low dissipation technology devices. Recent progress demonstrated the long-distance transport of spin signals across ferromagnetic insulators[1]. Antiferromagnetically ordered materials are however the most common class of magnetic materials with several crucial advantages over ferromagnetic systems. In contrast to the latter, antiferromagnets exhibit no net magnetic moment, which renders them stable and impervious to external fields. In addition, they can be operated at THz frequencies[2,3]. While fundamentally their properties bode well for spin transport, previous indirect observations indicate that spin transmission through antiferromagnets is limited to short distances of a few nanometers [4–8]. Here we demonstrate the long-distance, over tens of micrometers, propagation of spin currents through hematite ($\alpha$-$Fe_2O_3$)[9,10], the most common antiferromagnetic iron oxide, exploiting the spin Hall effect for spin injection. We control the spin current flow by the interfacial spin-bias[11] and by tuning the antiferromagnetic resonance frequency with an external magnetic field[11]. This simple antiferromagnetic insulator is shown to convey spin information parallel to the compensated moment (Néel order) over distances exceeding tens of micrometers. This newly-discovered mechanism transports spin as efficiently as the net magnetic moments in the best-suited complex ferromagnets[1]. Our results pave the way to ultra-fast, low-power antiferromagnet-insulator-based spin-logic devices that operate at room temperature and in the absence of magnetic fields.**

In both ferromagnetic and antiferromagnetic insulators (AFI), spin angular momentum can be transported by spin-wave excitations of the magnetic moments, called magnons. In easy axis antiferromagnets, the two degenerate magnon modes have left or right circular polarization, carrying finite, but opposite, angular momenta[12]. At thermal equilibrium, these degenerate magnons cannot be populated separately since their excitations frequencies are equal, leading to no net transport. In AFI/heavy metal (HM) bilayers, however, an interfacial spin-accumulation can generate an imbalance of the magnon population, enabling spin transport. Such a "spin-bias", along either the Néel vector $\boldsymbol{n}$ or the field induced magnetization $\boldsymbol{m}$, could be used to efficiently excite or annihilate the magnon modes depending on the bias sign. This potentially enables one to probe the antiferromagnetic spin conductance since the magnon modes independently transport spin angular momentum. In parallel, thermal excitations by Joule heating could propagate magnons, through the spin-Seebeck conductance, in the presence of a field induced net magnetization $\boldsymbol{m}$ [13,14]. Low magnetic dampings have been reported for AFIs [15,16], thus these materials naturally lend themselves to efficient long-distance spin transport experiments despite lack of observations[4–8].

In a recent report, signatures of long-distance spin-superfluid, rather than diffusive, transport through an antiferromagnet were claimed, based on only thermal transport, crucially without detecting the expected accompanying spin-injected transport signal. As discussed in various recent works[17–19], other contributions such as spatially extended thermal gradients can explain non-local thermal signals. By observing the first long-distance spin transport signal by spin-injection, we unambiguously demonstrate that it is consistent with diffusive rather than superfluid transport, accompanied by a thermal signal that decays over even longer distances (as discussed below and in the Supplemental[20]). Our results clearly rule out spin-superfluidity and demonstrate that long-distance spin transport through an antiferromagnet is possible even in the diffusive regime.

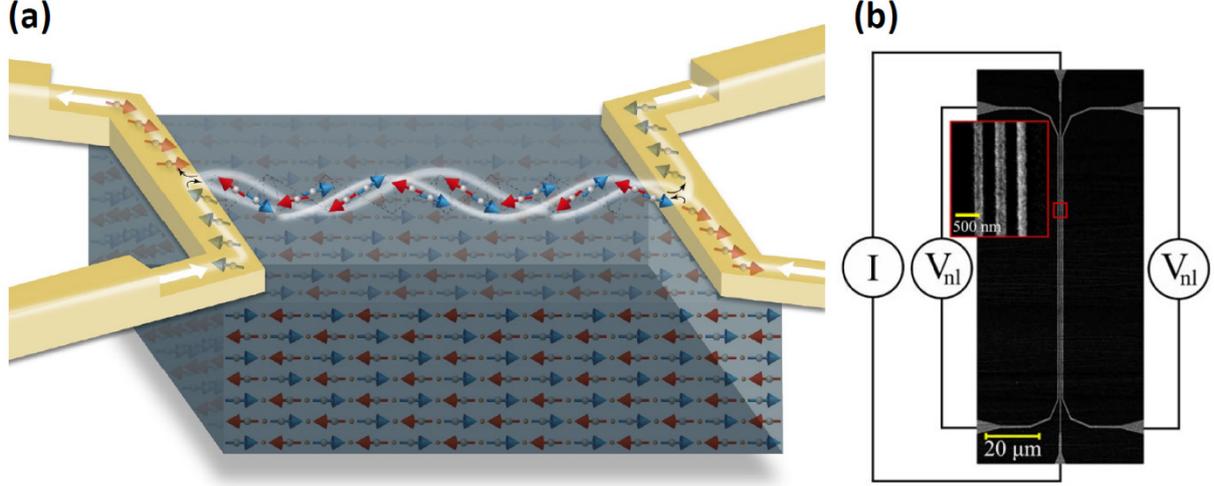

*Fig. 1 (a) **Schematic of the experiment**. Two Platinum (Pt) wires are deposited on the insulating easy axis antiferromagnet α-Fe$_2$O$_3$. A bias charge current through the left platinum wire generates an interfacial spin-accumulation at the Pt/α-Fe$_2$O$_3$ interface through the Spin-Hall effect. For a spin accumulation along the Néel vector $\boldsymbol{n} = (\boldsymbol{m_A} - \boldsymbol{m_B})/2$ as represented here, this spin-bias breaks the antiferromagnetic symmetry. The spin of conduction electrons in the platinum flip whilst scattering off the Pt/α-Fe$_2$O$_3$ interface and transfer their angular momentum into the antiferromagnet. The two antiferromagnetic magnon modes, which possess opposite circular polarization, are either generated or annihilated for opposite interfacial spin accumulations (whose directions are defined by the sign of the applied charge current). These excited magnons diffuse to the right platinum wire, where the reciprocal process occurs for magnon absorption, and the generated spin current is detected by the inverse spin-Hall effect. (b) **SEM image of a typical device with wire spacings of 200 nm and 250 nm.** The horizontal lines are platinum injector and detector wires, which are connected to chromium/gold contacts. Schematic current and voltage connections are indicated.*

To investigate the spin transport mechanisms in AFIs, we use a non-local geometry[1] with platinum (Pt) wires deposited on a micrometer thick sample of α-Fe$_2$O$_3$ (Fig. 1 (a)). A charge current, **I**, passes through a Pt wire, inducing two effects: (i) The spin-Hall effect (SHE) produces a transversal spin current, leading to a spin accumulation $\boldsymbol{\mu}$, at the Pt/α-Fe$_2$O$_3$ interface[21]. This accumulation may couple to the AFI order and thus generate a spin current which carries net angular momentum (see Fig. 1 (a)). (ii) Joule heating of the Pt wire induces a lateral temperature gradient $\Delta T$, which generates a spin-Seebeck induced thermal spin current. Ultimately, the total spin current, $V = V_{1\omega} + V_{2\omega}$, is a combination of both effects detected at a non-local Pt detector via the inverse SHE (ISHE)[21]. The even component, related to the spin-Seebeck conductance $\boldsymbol{S}$, can be determined from $V_{2\omega} = R_{2\omega} * [(+I)^2 + (-I)^2]/2$, and the odd component, describing the spin conductance $\boldsymbol{G}$, from $V_{1\omega} = R_{1\omega} * [(+I) - (-I)]/2$ to remove any thermal contributions[17]. The non-local signals $V_{2\omega}$ and $V_{1\omega}$ arise from spin currents carrying angular momentum along the antiferromagnetic Néel order $\boldsymbol{n} = (\boldsymbol{m_A} - \boldsymbol{m_B})/2$ and magnetic field **H** induced moment $= (\boldsymbol{m_A} + \boldsymbol{m_B})/2$. We therefore write the nonlocal resistances $R_{2\omega}$ and $R_{1\omega}$ using a phenomenological model based on two channel transport (see Supplemental[19]):

$$\begin{cases} R_{1\omega} = G_n(\boldsymbol{n} \cdot \boldsymbol{e_\perp})^2 + G_m(\boldsymbol{m} \cdot \boldsymbol{e_\perp})^2 & (Eq.\,1.a) \\ R_{2\omega} = S_n(\boldsymbol{n} \cdot \boldsymbol{e_\perp}) + S_m(\boldsymbol{m} \cdot \boldsymbol{e_\perp}) & (Eq.\,1.b) \end{cases}$$

where $\boldsymbol{e_\perp}$ is a unit vector normal to the Pt wires, i.e parallel to the current induced spin-accumulation μ. All four coefficients $(G_n, G_m, S_n, S_m)$ depend generally on the direction of $\boldsymbol{n}$ and $\boldsymbol{m}$. Finally, it should be noted that $R_{1\omega}$, and therefore $G_n$ and $G_m$, are direct measurements of the spin conductance of the antiferromagnet.

Our study requires full control of the Néel vector **n**, which cannot be achieved by aligning the magnetization with small external fields such as in ferromagnets. However, in easy-axis antiferromagnets with a low anisotropy like α-Fe$_2$O$_3$, one can control the direction of Néel vector with a field of a few Teslas. Above the spin-flop field H$_c$ (of about 6 T at 200K [10]), the Néel vector of α-Fe$_2$O$_3$ reorients in the entire the 500 μm thick sample perpendicular to the applied field **H** and we can control the antiferromagnetic order by sweeping an in-plane magnetic field along different directions (See Supplemental[20]). We then explore the spin transport signal in devices with injector–detector distances ranging from 200 nm to 80 μm (See Fig. 1 (b) and methods).

We first consider a device geometry with Pt wires oriented along **x**, the in-plane axis onto which the easy-axis is projected (see Supplemental[20]). Initially, the Néel vector **n** is approximately perpendicular to the spin accumulation, $\mu = \mu y$. As we sweep **H** along **x**, **n** rotates smoothly and becomes perpendicular to **H** (along **y**) when the field reaches $H_c$. In Fig. 2 (a) we find that the spin conductance signal $R_{1\omega}$ exhibits a maximum at $H_c$ (around 6 T) and remains non-zero at larger fields, whilst $R_{2\omega}$ is zero. For **H** > **H$_c$**, $R_{1\omega}$ is finite due to the reoriented Néel vector **n** along the spin-accumulation $\mu y$ (see sketch in Fig. 2 (a)). Therefore, we can identify the Néel spin conductance $G_n$ as the primary mechanism to carry angular momentum; the spin accumulation along **n** excites a distinct antiferromagnetic magnon mode, annihilating the other. This mechanism also naturally explains the sharp peak of $R_{1\omega}$ at $H_c$. Around $H_c$, the field compensates the anisotropy energy and the magnon gap of one mode strongly decreases[16]. A full gap closure theoretically leads to a divergence of $G_n$. Based on this explanation, we have theoretically modelled the experiment as shown in Fig. 2. We find that for our geometry, the magnon gap of one mode is reduced by a factor of 10 (see model in Supplemental[20]). This illustrates that $G_n$ depends on both the magnon gap and direction of **n**, and thus can be tailored. Additionally, we find that the Néel spin-Seebeck conductance $S_n$ is negligible, resulting in the absence of $R_{2\omega}$ below and above **H$_c$**.

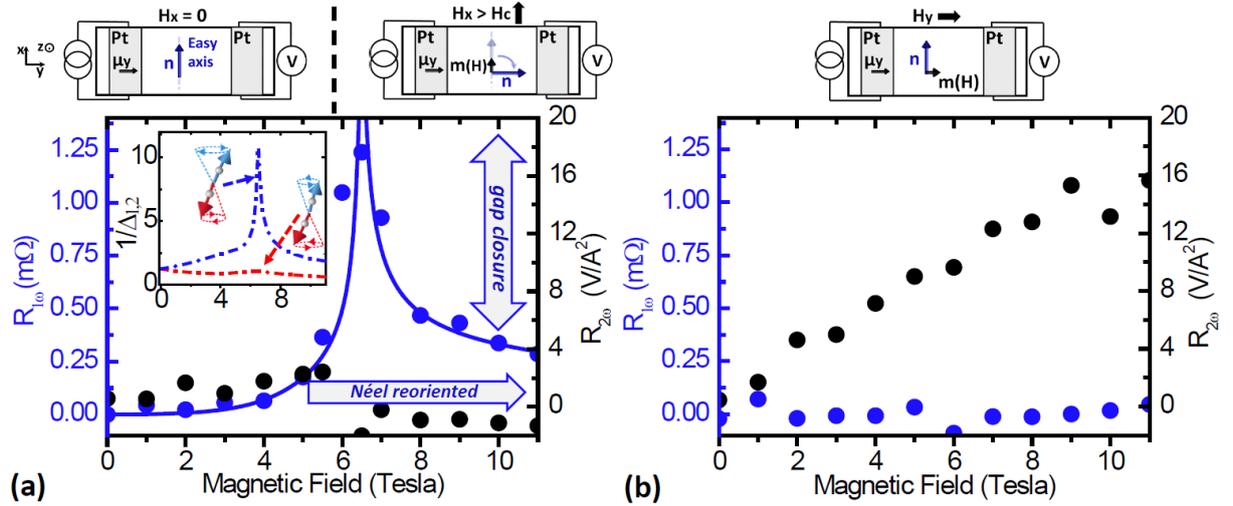

*Fig. 2.* ***Spin transport in the geometry with platinum wires along the x axis*** *(spin accumulation μ$_y$ along **y**): (a).* ***Magnetic field parallel to the Pt wires (along x)****: The non-local spin signal **R$_{1\omega}$** (blue dots) is zero at small fields, and finite for magnetic fields larger than the spin-flop field **H$_c$**, where the Néel vector is parallel to the spin accumulation μ$_y$. Around **H$_c$**, the spin transport signal reveals a sharp peak. The thermal spin-Seebeck signal **R$_{2\omega}$** shows low values (black dots).* ***Theoretical spin signal R$_{1\omega}$*** *(solid lines): The theoretical R$_{1\omega}$ is determined by the inverse of the lowest magnon gap Δ and by the projection of the spin-accumulation on the Néel vector (μ**y** · **n**). The transmitted signal is therefore maximum at **H$_c$** and remains non-zero for large magnon gaps at larger fields for which **n** is parallel to μ**x**. (Inset:* ***Theoretical inverse magnon gaps 1/Δ$_{1,2}$ of the two circularly polarized magnon modes****(dotted lines): The magnon gap of one mode is reduced by about a factor 10 at **H$_c$**, whilst the gap of the second one is enhanced. The dynamical opening angle of the two sub-lattices is exaggerated in the sketches.)* 
*(b)* ***Magnetic field perpendicular to the Pt wires (along y)****: The field induced magnetization **m** is parallel to μ**y**, and **n** remains along x. Here, R$_{1\omega}$ remains equal to zero (blue dots). In parallel, the R$_{2\omega}$ increases linearly with the applied magnetic field, indicating a large magnetic Spin-Seebeck conductance **S$_m$**.*

We then study a second field direction where the Pt wires are still oriented along **x** but the magnetic field **H** is applied along **y** (Fig. 2 (b)). For this geometry, there exists a field induced magnetization **m** parallel to the spin accumulation μ**y** as seen in Fig. 2 (b). The absence of $R_{1\omega}$ is in line with the prediction of our model that the magnetic spin conductance $G_m$ should be reduced by a factor Tχ ≪ 1 as compared to $G_n$ (with T the temperature and χ the susceptibility)[11]. However, the spin-Seebeck signal $R_{2\omega}$ now contributes, increasing linearly with **H**. The spin-Seebeck conductance $S$ is non-zero only in the presence of a field induced magnetization **m** along μ, indicating that $S_m \gg S_n$. In antiferromagnets, **H** does not break the symmetry of the Néel vector **n** and the spin-accumulation is only along **m**, hence, $S_n$ vanishes in the absence of sublattice-symmetry breaking.

We can initially conclude that we can generate a spin-current that propagates through an AFI, mediated by the Néel spin conductance $G_n$ and the magnetic spin-Seebeck conductance $S_m$ (full angular scans with theoretical fits in the Supplemental[20]).

For devices, the application of strong fields is cumbersome. A possible device geometry field-free transport has the platinum wires along **y**, i.e perpendicular to the easy-axis at zero applied field, as shown in Fig. 3 (a-b).

We first analyze this geometry for **H** applied along **x**. As seen in Fig. 3 (a), the spin-Seebeck signal $R_{2\omega}$ increases only above $H_c$, when **n** reorients perpendicular to **H** and a sizeable field induced moment **m** along μ**x** appears.

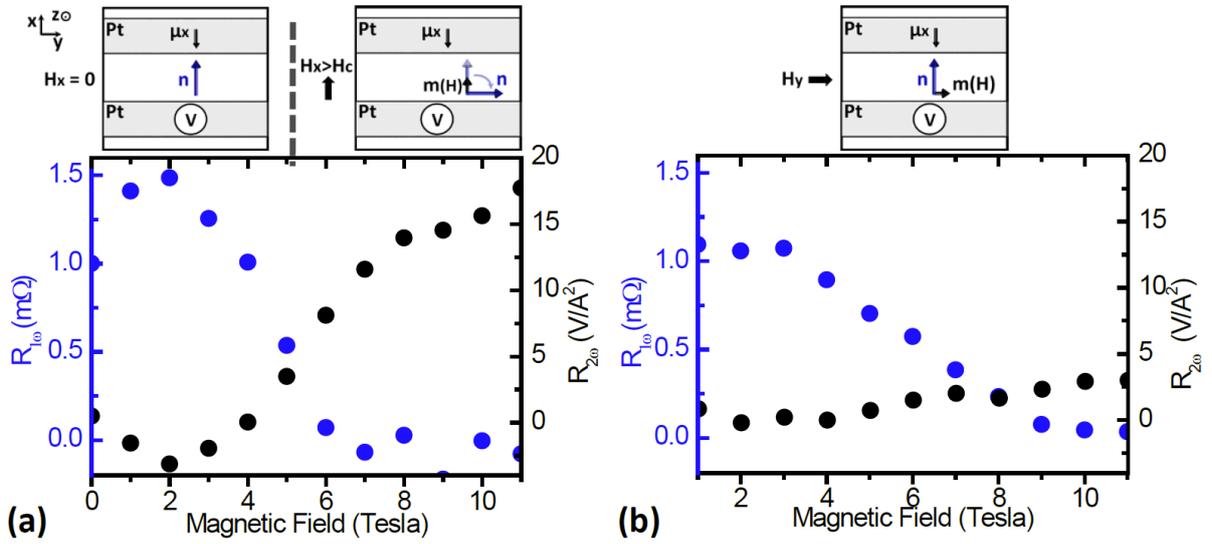

*Fig. 3. **Spin transport in the geometry with platinum wires along the y axis** (spin accumulation **μ** along x). (a). **Magnetic field perpendicular to the Pt wires (along x)**: The spin signal $R_{1\omega}$ is finite at zero and small applied magnetic fields and sharply decreases at the spin-flop when the Néel vector **n** aligns along x, i.e perpendicular to the spin accumulation μx. Conversely, the spin-Seebeck signal $R_{2\omega}$ is initially zero and shows an enhancement at the spin-flop field, when a field induced magnetization **m** emerges along the spin accumulation μx. (b). **Magnetic field parallel to the Pt wires (along y)**: The spin signal $R_{1\omega}$ decreases steadily with the applied magnetic field due to reorientation of the Néel order out of the (xy) plane at large field (see Supplemental[20]). The spin-Seebeck signal $R_{2\omega}$ is equal to zero as the field induced magnetization **m** is always directed perpendicular to the spin accumulation μx.*

The striking feature in this geometry is the presence of a strong spin signal $R_{1\omega}$ at zero applied field due to the easy axis orientation where the Néel vector **n** is already parallel to μx. This observation confirms the high field measurements in the previous geometry: a large Néel spin conductance $G_n$ exists under spin bias, even without a strong reduction of the magnon gap. This field-free spin-transport demonstrates the potential of antiferromagnets for use in applications. For **H** along **x** (see Fig. 3(a)), the sharp drop of $R_{1\omega}$ at larger fields reflects the reorientation of the Néel order **n** along the wire (y-direction), perpendicular to μx.

Then, we apply the field along **y**. In Fig. 3 (b), the Spin-Seebeck signal $R_{2\omega}$ remains zero as the field is applied parallel to the platinum wires, whilst the spin signal $R_{1\omega}$ is again finite as the Néel order is insensitive to small fields. Thus, the spin signal reveals strongly distinct field dependences for the two device geometries. This shows that active control of the Néel order direction is key for spin transport, which is likely to be problematic in multi-domain samples and thus can explain low efficiency spin transport in previous studies (in our 500 μm thick samples large domains are found, see Supplemental[20]). We note that the amplitude of the signal at small fields is comparable to the signal at the reduced magnon gap, possibly arising from different interface transmissivities[20]. Another intrinsic contribution most probably comes from spin-relaxation processes depending on applied fields; at $H_c$, the changes of the magnon dispersion curve could lead to different dissipation channels.

Finally, we turn to the experimental determination of the spin diffusion lengths. This is a key point to determine the spin transport regime, for which many different predictions have been made[11,22,23]. We focus in Fig. 4 only on the distance dependence of the spin signal $R_{1\omega}$, as the thermal heating associated with $R_{2\omega}$ is not localized at the position of the injector[18,24]. In Fig. 4, we find spin-transport over tens of μm with a linear decay up to a few μm. Here we want to stress that this feature (and the $R_{2\omega}$ signal up to 80 micrometers, see Supplemental[20]) might lead one to conclude a spin-superfluid regime as very recently discussed[25] at low temperatures, even though one certainly observes here diffusive transport. Only for distances larger than the spin diffusion lengths is an exponential amplitude decay predicating diffusive transport is observed[1]. Moreover, $R_{1\omega}$ is, in both configurations, linear with the bias current and presents no threshold up to 2.5 x $10^{11}$ A/m$^2$, which would exist due to anisotropy effects in the spin-superfluid regime. So our observations clearly identify diffusive spin transport[11,22,23], as expected at our relatively high temperature of 200K.

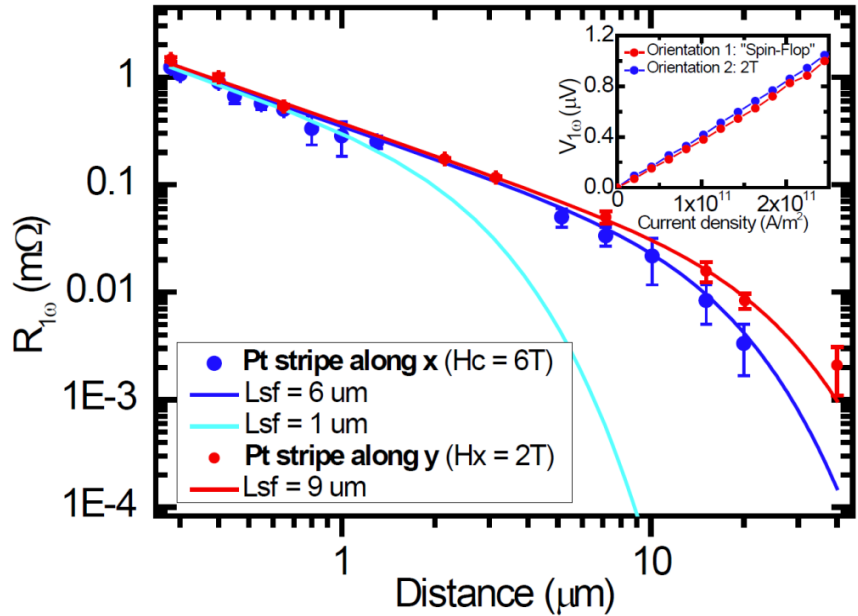

*Fig. 4* **Distance dependence of the spin signal $R_{1\omega}$ at 200K** *(blue dots). A spin-transport signal persists for distances of tens of micrometers at both the spin-flop field (Platinum wires oriented along **x**, blue dots) and small applied fields (Platinum wires oriented along **y**, red dots). Only the spin signal is shown as its source is precisely localized at the Pt injector position (for the spin-Seebeck signal, the signal persists for distances of more than 80 μm. However conclusions are not easy to draw as the heating and thus spin current source cannot be considered as localized in a single spot for this case[24], see also Supplemental[20]). The fitted solid lines are obtained from a one dimensional spin diffusion equation[1]. For distances smaller than the spin-diffusion length, the amplitude decay is linear with the distance between the platinum wires and exponential above the spin-diffusion length. (Inset) The spin-bias voltages, $V_{1\omega}$, are linear for both configurations in accordance with a non-equilibrium spin-transport mechanism.*

We determine the spin diffusion lengths to be 6 ± 1 μm and 9 ± 2 μm at the spin-flop field and small-applied fields for each geometry, more than two orders of magnitude larger than reports using AFI/FM multilayer thin films [5–7,26]. In these systems, magnetic correlations between FM and AFI grains are crucial with the presence of the direct and spin-flop couplings[27]. As shown above, a Néel order **n** parallel to **μ** is crucial to get an efficient spin-transport. A second key issue in many AFI thin films is the presence of multi-domains affecting the mode polarization[5,27]. Our observed micrometer spin-diffusion lengths are moreover in accordance with theory works[28] and rely on the low Gilbert damping of α-Fe$_2$O$_3$ (α < $10^{-4}$ [29]), although the question of antiferromagnetic spin-wave relaxation processes remains an open debate (with ultra-small linewidths of $10^{-5}$ T[30], constant with field[31], and multi-magnon scattering[32]). This is illustrated here as the spin diffusion length is slightly smaller at $\mathbf{H_c}$ (9 ± 2 μm compared to 6 ± 1 μm), indicating that the magnon gap reduction at the spin-flop field comes at the cost of other dissipation channels . Investigations of AFIs with lower damping would reveal the virtually unexplored relaxation processes of antiferromagnetic spin waves and open possible THz magnon spin transport over macroscopic distances, a path, which is just starting to be explored by our work.

**Methods**

**Lithography:**

The non-local measurements were carried out based on a sample geometry that was defined using electron beam lithography and the subsequent deposition and lift-off of a 7 nm platinum layer by DC sputtering in an argon atmosphere at a pressure of 0.01 mbar. The non-local wires were contacted using a bilayer of chromium (6 nm) and gold (32 nm). The geometry consisted of three wires 80 µm in length (L) and 350 nm wide in an asymmetric layout with wires patterned both parallel and perpendicular to the easy axis of the sample. The center wire was used to carry the injection current of 300 µA whilst non-local voltages were measured in the left and right wires and normalized to a non-local resistance. The separations of the left-center and right-center wire configurations differed and ranged from 200 nm – 80 µm. The sample was mounted to a piezo-rotating element in a variable temperature insert that was installed in a superconducting magnet capable of fields up to 12 T and cooled with liquid helium. The temperature was fixed at 200 K, far below the Morin transition to guarantee the sample was in the antiferromagnetic easy axis phase, and the field was swept in-plane either parallel or perpendicular to the easy axis of the α-$Fe_2O_3$. For the rotation measurements, the sample was rotated in a constant field.

For the center-to-center separations of more than 10 µm, a second geometry was utilized to limit the geometric impact on the signal, allowing us to continue to approximate the length of the wire to be far greater than the separation. For these measurements, the wire length was increased to 160 µm (2L) whilst the width remained constant. The injection current was also increased to 600 µA to increase signal to noise. As can be seen in the inset of figure 4 of the main text, the signal is linear with current up to at least 600 µA. To allow for accurate scaling to account for the increase in signal from the doubled current and doubled length, calibration distances of 500 nm were added in this second geometry where the scaled signal was found to be consistent with comparable distances in geometry one.

**Authors contribution:**
R.L. and M.K proposed and supervised the project. R.L performed the experiments with A.R and technical support from J.C and L.B. A.R patterned the samples. R.L, S-A.B, A. R. analysed the data with inputs from M.K, A.B. and R-A. D. S-A. B. performed the theoretical calculations with the assistance of A.Q., A.B and R-A.D. R.L wrote the paper with S.B, A.R and M.K. All authors commented on the manuscript.

Correspondence and requests for materials and information on Methods, Supplemental Notes should be addressed to R.L. or M.K (e-mails: rolebrun@uni-mainz.de or klaeui@uni-mainz.de).